# Control of chiral orbital currents in a colossal magnetoresistance material

Yu Zhang[1], Yifei Ni[1], Hengdi Zhao[1], Sami Hakani[2], Feng Ye[3], Lance DeLong[4], Itamar Kimchi[2*]

and Gang Cao[1*]

[1]*Department of Physics, University of Colorado at Boulder, Boulder, CO 80309, USA*

[2]*School of Physics, Georgia Institute of Technology, Atlanta, GA 30332, USA*

[3]*Neutron Scattering Division, Oak Ridge National Lab, Oak Ridge, TN 37831, USA*

[4]*Department of Physics and Astronomy, University of Kentucky, Lexington KY 40506, USA*

***Abstract*** Colossal magnetoresistance (CMR) is an extraordinary enhancement of the electric conductivity in the presence of a magnetic field. It is conventionally associated with a field-induced spin polarization, which drastically reduces spin scattering and thus electric resistance. However, ferrimagnetic $Mn_3Si_2Te_6$ is an intriguing exception to this rule: it exhibits a 7-order-of-magnitude reduction in *ab*-plane resistivity with a 13-Tesla anisotropy field which occur only when a magnetic polarization is avoided [1]. Here we report an exotic quantum state that is driven by *ab*-plane chiral orbital currents (COC) flowing along edges of $MnTe_6$ octahedra. The *c*-axis orbital moments of *ab*-plane COC couple to the ferrimagnetic Mn spins to drastically increase the *ab*-plane conductivity (CMR) when an external magnetic field is aligned along the magnetic hard *c*-axis. Both the COC state and its CMR are extraordinarily susceptible to small DC currents exceeding a critical threshold, and a hallmark of this COC state is an exotic time-dependent, bistable switching mimicking a first-order "melting" transition. The control of the COC-enabled CMR and bistable switching offers a fundamentally new paradigm for quantum technologies.

* *gang.cao@colorado.edu*; *ikimchi3@gatech.edu*

***Introduction*** A necessary characteristic of all known colossal magnetoresistance (CMR) materials is an alignment of magnetic spins that drastically reduces electron scattering, thus electric resistance [2-10]. However, ferrimagnetic $Mn_3Si_2Te_6$ is an intriguing exception to this rule, in that the *ab*-plane electric resistivity is reduced by up to 7 orders of magnitude only when a magnetic field H is applied along the magnetic hard *c*-axis or when a saturated magnetic state is absent [1]. In contrast, the *ab*-plane resistivity decreases by only 20% at most when the magnetization is saturated by H aligned with the magnetic easy *a*-axis (or *ab*-plane) (**Figs.1a-1b**). The data in **Fig. 1b** also reveal a large anisotropy field of at least 13 T. This observation is equally intriguing because magnetic anisotropy is usually a result of spin-orbit coupling; but here the orbital momentum is zero for the $Mn^{2+}$ ($3d^5$) ion with a half-filled 3*d* orbital. This behavior signals a new type of CMR, which has since been confirmed [11], with a peculiar anisotropy that defies existing precedents [2-10] and indicates a novel quantum state that has yet to be identified and understood.

Herein we report extensive evidence of a novel quantum state in $Mn_3Si_2Te_6$: The new state is defined by its unconventional CMR, anisotropy and extraordinary response to application of small DC currents I, including a first-order transition with bistable switching. The switching consists of an abrupt jump in voltage V which takes seconds or minutes to occur after application of a small current to the sample without further stimulus. This time delay and voltage jump drastically increase when H is oriented along the magnetic hard *c*-axis where the CMR occurs [1] but are otherwise absent. Applying H || *c*-axis also induces a surprising DC tunneling behavior that features $\Delta V/\Delta I = 0$. In short, this state becomes highly conducting (CMR) and resilient when H || *c*-axis is employed but is insulating and "melts" (mimicking an ice-to-water phase transition) via first-order bistable switching when a small DC current exceeding a certain threshold is applied.

Each of the key phenomena, i.e., the CMR, magnetic anisotropy, and bistable switching, defies conventional wisdom and models; the *simultaneous* occurrence of all these phenomena indicates that a novel paradigm is required to understand the physics of the underlying state. Here we argue that these phenomena can be explained in terms of a state of intra-unit-cell, *ab*-plane chiral orbital currents (COC) that generate net c-axis orbital magnetic moments ($M_{COC}$) which couple with the simultaneously ferrimagnetically ordered Mn spins ($M_{Mn}$) (**Figs. 1e-1f)**. Below the Curie temperature $T_C$ = 78 K, the *ab*-plane COC circulate along Te-Te edges of the $MnTe_6$ octahedra, thereby producing net $M_{COC}$ primarily oriented along the *c*-axis (**Fig. 1e**). $M_{COC}$ arising from Te-orbitals are necessarily coupled with $M_{Mn}$, giving rise to an unusual spin-orbit coupling that explains the observed large anisotropy field (**Fig. 1b**). Therefore, the observed magnetization results from local $M_{Mn}$ and $M_{COC}$. Application of H || *c*-axis can thereby amplify the *ab*-plane COC and $M_{COC}$ and underpins the CMR (**Fig. 1a**). The COC state, $\Psi_C$, is exceedingly sensitive to application of small DC currents I and converts to a trivial state, $\Psi_T$, via first-order, bistable switching when I > 2 mA. The rigid coupling of the COC to the Te sublattice and $M_{Mn}$ causes $\Psi_C$ to remain metastable over long time scales set by Te atomic motion (as opposed to only electrons or Mn spins which respond on picosecond time scales [12]). Consequently, the bistable switching or melting of $\Psi_C$ requires *seconds or minutes* to occur after initial application of a small I to the sample. Our observed temperature-current-field (T-I-H) phase diagram (**Fig.1g**) shows that $\Psi_C$ exists under a sheet in T-I-H space that expands with increasing H || *c*-axis.

A COC state was initially proposed and investigated in studies of high-$T_C$ cuprates [13-20] and was later invoked in investigations of other materials, such as iridates [21-23] and Kagome superconductors [24, 25]. A more recent study reports a transport signature of loop current in a Kagome metal [26]; the data show the signal in a charge-ordered phase of the material, which is attributed a loop-current phase with spontaneously broken mirror symmetries [26]. It is increasingly clear that COC may be widespread beyond the forementioned materials [18, 24-27]. However, no macroscopic transport phenomena attributed to COC that coexist with a long-range magnetic order such as those observed in this work have ever been reported. The signature of COC coupled with long-range magnetic order can be particularly subtle because mirror and time reversal symmetries are already broken, but effects of COC can be exotic and strong. Indeed, the observed CMR [1, 11], magnetic anisotropy and time-dependent bistable switching are both rare and striking. In the following we present the first-such experimental evidence, and then discuss how COC can adequately account for the observations [28].

***Crystal and magnetic structures*** $Mn_3Si_2Te_6$ crystallizes in a trigonal space group *P-31c* (No. 163) with two inequivalent Mn sites, Mn1 and Mn2 [29], forming a *trimer-honeycomb lattice* (**Figs.1e-1f**). It orders ferrimagnetically at $T_C$ = 78 K [29, 30] with the Mn spins antiferromagnetically coupled along the *c*-axis [1]. Recent neutron diffraction data reveal a noncollinear magnetic structure below $T_C$, in which the Mn spins lie predominantly within the *ab*-plane, but tilt both within the *ab*-plane and toward the *c*-axis by 10$^o$ under ambient conditions (**Fig. 1d**) [31]. Application of H || *c*-axis causes a *gradual* tilting of the spins toward the *c*-axis while retaining the underlying antiferromagnetic configuration along the *c*-axis [31], which reinforces a key point that any models based on the Mn spins alone cannot explain the CMR that is also current sensitive.

***Structural, transport, and magnetic properties in magnetic fields*** The *a*-axis resistivity $\rho_a$ and the *c*-axis resistivity $\rho_c$ both rise rapidly below $T_C$, reaching values as high as $10^7$ Ω-cm (**Fig. 1a**) [1]. The *a*-axis magnetization $M_a$ readily saturates to a value $M_s$ = 1.56 $\mu_B$/Mn for $\mu_oH_{||a}$ > 0.05 T, indicating the magnetic easy *a*-axis (red dashed line in **Fig. 1b**). In contrast, the *c*-axis magnetization $M_c$ cannot fully attain the *a*-axis value, even for $\mu_oH \geq 13$ T (blue dashed line in **Fig. 1b**) [1]. Such a large anisotropy field (> 13 T) is conventionally unexpected because the spin-orbit coupling for the $Mn^{2+}$ ($3d^5$) ions is expected to be negligible, which suggests a novel type of spin-orbit coupling may be at play in $Mn_3Si_2Te_6$. Moreover, the measured average magnetic moments for Mn1 and Mn2 are 4.55 and 4.20 $\mu_B$, respectively [31], which are significantly smaller than 5 $\mu_B$ anticipated for $Mn^{2+}$ ($3d^5$) and suggest that the measured moments may result from partial cancellation between $M_{Mn}$ and $M_{COC}$. Accordingly, an upper limit of $M_{COC}$ could be on the order of 0.1 $\mu_B$ [28].

Further clues can be gained from the observed magnetostriction data that indicate the *a*-axis undergoes a significant expansion $\Delta a/a$ when H || *c*-axis but remains essentially unchanged when H || *a*-axis (**Fig. 1c**), which indicates a strong, anisotropic coupling of H to the lattice. In other words, increasing H || *c*-axis favors an increase of orbital area and a strong *c*-axis magnetoelastic coupling. The magnetostriction is relatively easy to associate with orbital moments of a COC state because the Lorentz force acts to expand an *ab*-plane loop current circulating on the rigid Te-Te edges of $MnTe_6$ octahedra when H || *c*-axis. The net orbital magnetization, which equals the loop current times the orbital area, naturally increases due to increasing the orbital area. It is also noteworthy that Ge doping that enhances the CMR also expands the *ab*-plane, whereas Se doping that weakens the CMR shrinks the *ab*-plane [28]. We conclude a strong, anisotropic magnetoelastic coupling is consistent with the presence of a COC state, as we further discuss below.

***Transport and magnetic properties with small DC currents*** $\rho_a$ at H = 0 is reduced by up to 6 orders of magnitude when applied current I is increased from 10 nA to 10 mA at low temperatures (**Fig. 2a**). $T_C$ rapidly decreases with increasing I || *a*-axis (and *c*-axis [28]) (**Fig. 2b**). Specifically, $T_C$ decreases from 83 K at 1μA to 22 K at 1.8 mA, and eventually vanishes at 2 mA (**Fig. 2b**). $M_c$ behaves similarly with increasing I || *a*-axis (**Fig. 2c**). Clearly, $T_C$ evolves into a first-order

transition with increasing I, until completely suppressed for I $\geq$ 2 mA. However, the magnitude of $\rho_a$ in the vicinity of $T_C$ remains essentially unchanged (**Fig. 2b)**. This behavior suggests that small currents (< 2mA) hardly affect the spin scattering of electrons near $T_C$ but effectively weaken the spin exchange coupling that drives $T_C$. (This current-controlled behavior is vastly different from that seen in other systems [32, 33].) In contrast, $T_C$ systematically shifts to higher temperatures with increasing H || *c* axis (**Fig. 1a**).

***Transport properties in magnetic fields and with small DC currents*** The response of $\rho_a$ to I drastically changes when H **||** *c*-axis (**Fig. 2d)**. H || *c*-axis recovers the magnetic order otherwise suppressed by I, and simultaneously reduces $\rho_a$, leading to a metallic state below $T_C$ for I $\leq$ 2 mA. However, for I > 2 mA, H only slightly reduces $\rho_a$, leaving a much weaker conducting state; the CMR is thus no longer present for I = 5 and 10 mA, despite $\mu_o H_{||c}$ = 14 T.

Combining the data for $\rho_a$(H=0, I) in **Fig. 2a** and $\rho_a(H_{||c}$=14T, I) in **Fig. 2d** yields the magnetoresistance ratio as a function of I, [$\rho_a$(14T, I) - $\rho_a$(0, I)]/$\rho_a$(0, I) at 10 K in **Fig. 2e**. Clearly, the CMR occurs only when I $\leq$ 2 mA and abruptly disappears when I > 2 mA, which closely tracks the current dependence of $T_C$ in **Fig. 2f**. While H **||** *c*-axis enables the CMR, applied current I > 2 mA weakens and eventually suppresses the ferrimagnetic state, thereby recovering $\Psi_T$ without the CMR. The correspondence exhibited by the transport anomalies is even more evident in the behavior of the I-V characteristics shown below.

***The anomalous I-V characteristic*** The I-V characteristic for H = 0 exhibits a first-order transition characterized by a critical current $I_C$ that separates the two distinct states. In addition, there are two onsets of S-shaped negative differential resistance (NDR) [34, 35], $I_{NDR1}$ (< $I_C$) and $I_{NDR2}$ (>$I_C$) (**Figs. 3a-3b**). The I-V curve bends over at an onset of the first NDR regime at I = $I_{NDR1}$, above which an increase in I leads to a decrease in V. This first NDR1 regime ends at $I_C$ = 1.90 mA, where an abrupt increase in V occurs, indicating a first-order transition (**Fig. 3b**). This transition at $I_C$ is followed by the second onset of NDR at I = $I_{NDR2}$ [28].

The magnetic field dependence of the I-V characteristic illustrates a novel type of bistable switching (**Fig. 3c**). Upon the vanishing of $I_{NDR1}$ at $\mu_o H_{||c}$ > 3 T, an extraordinary region emerges with ΔV/ΔI = 0 for 0 ≤ I ≤ $I_C$ (**Fig. 3d).** Note that at 7 T and 14 T, V ≈ 0 as I increases from zero to $I_C$, a behavior strikingly similar to the DC Josephson effect [36] although no superconducting state is involved here (**Figs. 3c-3d**). This region disappears via an abrupt transition at $I_C$, and a more resistive state emerges at $I_{NDR2}$ (**Figs. 3d**). The phase diagrams in **Figs. 3e-3f** show the first-order transition at $I_C$ as a function of temperature or magnetic field, respectively. Note that $I_C$ behaves entirely differently when H || *a*-axis [28].

***Time-dependent bistable switching*** A particularly striking and unique feature of this COC state is that the first-order, bistable switching requires a finite time (e.g., seconds or minutes) and occurs without additional stimulus. The *a*-axis voltage $V_a$, measured as a function of time t at 10 K, is shown in **Fig. 4**. Each measurement of $V_a$ starts at t ≡ 0 when a constant I is applied and continued until t = 180 seconds (and 1,800 seconds [28]) has elapsed. For example, when I ≤ 2.030 mA at t = 0, $V_a$ decreases slightly near t = 1 second, and then remains constant for the rest of the time of measurements. A current increase of merely 0.25% (i.e., from 2.030 to 2.035 mA) causes $V_a$ to abruptly spike at t = 9.6 seconds, with a voltage increase of $\Delta V_a$ = 0.52 V, or $\Delta V_a/V_a$ = 30%, and then immediately stabilizes at a constant value (**Fig. 4a**). Further, slight increases of I progressively shorten the delay time for the switching from 9.6 seconds to 1 second. However, the change in $V_a$ or $\Delta V_a$ remains essentially the same, at 0.52 V. This protocol indicates that only two discrete states exist, separated by a first-order switching, which persists up to $T_C$ [28].

The switching process takes even longer when H || *c*-axis (**Fig. 4b**). The switching takes up to 114 seconds to occur at a significantly stronger I = 3.98 mA with $\mu_o H_{||c}$ = 7 T, but the switching leads to a much larger voltage increase, $\Delta V_a \simeq 0.99$ or $\Delta V_a/V_a \simeq 2000$ % (**Fig. 4b)**.

It is remarkable that the abrupt bistable switching completely vanishes when H || *a*-axis (**Fig. 4c**). The abrupt switching is instead replaced by a gradual, continuous change in $V_a$ with I. The stark contrast between data in **Fig.4b** for H **||** *c*-axis and **Fig.4c** for H || *a*-axis reinforces the following key points. Application of H || *c*-axis apparently strengthens $\Psi_C$ via the *c*-axis $M_{COC}$, which interlock with the Te sublattice and $M_{Mn}$ (**Fig. 4e**). It therefore takes a stronger I and a longer t for the charge carriers to alter the current configuration. The fact that this switching is always achieved via a first-order transition further emphasizes the key differences in current paths and orbitals involved in the two states. Conversely, application of H || *a*-axis, where the CMR is absent, tilts the *c*-axis $M_{COC}$ towards the *a*-axis, which weakens or even destroys $\Psi_C$ (**Fig. 4f)**. The continuous increase in $V_a$ with increasing I for H **||** *a*-axis suggests that $\Psi_C$ is no longer robust or distinct at t = 0. That the data in **Fig. 4c** still bear some resemblance to the data in **Fig. 4b** may imply the existence of a "mixed state" consisting of coexisting $\Psi_C$ and $\Psi_T$ when H || *a*-axis.

Our extensive measurements further reveal that the critical current for switching is an intrinsic parameter; that is, no switching will happen for any smaller current, no matter how close it is to the critical current and how much time elapses (e.g., 30 minutes [28]). Moreover, $\Delta V_a/V_a$ as a function of H || *c*-axis is unchanged for all temperatures below 70 K, which demands a bistable state.

***Absence of Joule heating*** Self-heating effects cause a continuous drift in local temperature. They are generally isotropic or diffusive and vary continuously with changing current. Such behavior is ruled out in the present study by (i) the abrupt nature of the switching without any drifting, (ii) the independence of the two discrete values of $V_a$ on the magnitude of I in **Figs. 4a-4b**, (iii) the extremely anisotropic behavior in **Figs. 4b-4c**, and moreover, (iv) the electric energy W = $V_a$ x I x t dissipated by the applied current at each switching, for H = 0 and $\mu_o H_{||c}$ = 7 T, *decreases rapidly with increasing I*, which is clearly inconsistent with Joule heating (**Fig. 4d**).

***Discussion of the COC state*** The COC, which couple to the ferrimagnetic state, originate from spontaneous breaking of time-reversal symmetry below $T_C$ = 78 K and circulate along Te-Te edges of the $MnTe_6$ octahedra, producing $M_{COC}$ primarily oriented along the *c*-axis [28]. (Note that similar orbital moments of loop currents are discussed in the cuprates [19, 37]). In the absence of H || *c*-axis, the net circulation of the COC is zero. This is because the COC are allowed to circulate both clockwise and counterclockwise (**Figs.1e, 1f** and **Fig.5**), resulting in domains with two opposite directions of the COC. The disordered domains lead to strong scattering and high resistance. Application of H || *c* axis favors only one circulation (i.e., either clockwise or counterclockwise) and enlarges its domains (and concurrently reducing the domains with the opposite circulation). Since the overall conductivity is a sensitive function of the configuration of the COC domains, the alignment of the dominant circulation domains is crucial for the enhancement of $\Psi_C$ by H || *c* axis, and the sharp reduction in electron scattering evident in the $10^7$-CMR (**Figs.1a-1b**) [28]. Naturally, application of H || *a*-axis weakens the *c*-axis $M_{COC}$, thus $\Psi_C$, and necessarily causes suppression of the CMR.

Applying nonequilibrium I > 2mA weakens the order parameter of $\Psi_C$, therefore destabilizes $\Psi_C$ in favor of $\Psi_T$. This is expected since the COC configuration does not permit any net current because each of the four independent parameters that specify $\Psi_C$ [28] forms a circulating current within the unit cell with no net uniform component (**Figs. 5a-5c**). The resistivity is reduced with increased I, but at the same time, the increased I also weakens and eventually suppresses the ferrimagnetic state, and consequently $\Psi_C$, leading to a first-order phase transition that destroys $\Psi_C$ (**Figs. 2a-2d**).

The rigid coupling of the COC (and $M_{COC}$) to the Te sublattice and $M_{Mn}$ is a signature of $\Psi_C$. The COC extend over multiple atomic sites, with couplings that depend on the ionic positions and dictate the current strength. Therefore, $\Psi_C$ entails a particularly strong coupling between the lattice, the *c*-axis $M_{COC}$ and $M_{Mn}$ as well as the anisotropic magnetoelastic coupling. Such a rigid coupling in $\Psi_C$ explains the current-driven first-order transition from $\Psi_C$ to $\Psi_T$ (**Figs. 1-4**). This coupling also enables $\Psi_C$ to remain metastable over long time scales that are set by Te atomic motion and bond lengths or fluctuations/phonon effects rather than by electrons and/or Mn spins alone (**Fig.4a**). Application of H || *c*-axis, via the *c*-axis $M_{COC}$, strengthens the coupling, further extending the time scale (**Fig.4b)** but application of H || *a*-axis weakens or destroys the coupling (**Fig.4c**).

Our observations implicate an exotic quantum state of matter that yields unusual responses to small applied electric currents and magnetic fields. While further work, such as optical control of this state [38], will reveal more insights into the COC, the observed phenomena certainly promise wide-ranging opportunities for fundamental advances in condensed matter physics and future device applications.

**Important Notes:** The final version of this paper, which is more updated than this version, is to be published in *Nature*. The updated Methods, Extended Data Figures and updated references are presented in the final version of the paper.

***Figure Captions***

**Fig. 1. Physical properties in applied magnetic fields and phase diagram (a)** The temperature dependence of the *a*-axis resistivity $\rho_a$ at various magnetic fields. **(b)** The magnetic field dependence of the *a*-axis magnetoresistance ratio $[\rho_a(H)-\rho_a(0)]/\rho_a(0)$ and the magnetization M (dashed lines, right scale) for H || *c*-axis (blue) and H || *a*-axis (red). **(c)** The *a*-axis magnetostriction $\Delta a/a$ for H || *c*-axis (blue) and H || *a*-axis (red). **(d)** The crystal and magnetic structure [31]; the yellow arrows indicate the *ab*-plane expansion when H || *c*-axis denoted by the blue arrow. **(e)** the COC and corresponding *c*-axis $M_{COC}$; different colors indicate different magnitudes of $M_{COC}$ (see details in Fig.5 and Ref. 28). **(f)** The *ab*-plane view of the COC. **(g)** A T-I-H phase diagram. Note that H || *c*-axis and I || *a*-axis. $\Psi_C$ exists under the purple sheet in the T-I-H space and beyond it is $\Psi_T$.

**Fig. 2. Response of physical properties to DC currents and applied magnetic fields: (a)** The *a*-axis resistivity $\rho_a$ at various applied DC currents I for H = 0. **(b)** Enlarged plot of $\rho_a$ in the outlined region in (a). The dashed gray lines highlight the near-constant values of $\rho_a$ at and just below $T_C$. **(c)** The *c*-axis magnetization $M_c$ at various DC currents I along the *a*-axis for $\mu_o H$ = 0.5 T. Note the first-order transition evident in the $\rho_a$ and $M_c$ data induced by I = 1.5 and 1.8 mA.

**(d)** Temperature dependence of $\rho_a$ at various applied DC currents I for $\mu_o H = 14$ T. Note that the CMR is absent for I = 5 mA and 10 mA. **(e)** The change of $\rho_a$ for $\mu_o H = 14$ T and various values of the applied I. **(f)** $T_C$ at various values of I (generated from data in Figs. 2b-2c). Note that both the CMR and $T_C$ vanish for I > 2 mA.

**Fig. 3. The *a*-axis I-V characteristic (a)** At various temperatures and H = 0; **(b)** Details of the outlined area in (a); note the first-order transition at $I_C$. Note that the sudden voltage increase, $\Delta V = V_{NDR2} - V_C = 0.52$ V, results from a tiny current increase of 0.1 mA from the value of $I_C$. **(c)** I-V characteristics for T = 10 K and H || *c*-axis. **(d)** Details of the outlined area in (c). Note the regime where $\Delta V/\Delta I = 0$ emerges for $\mu_o H \geq 3$ T, and where $V_a \approx 0$ for $I < I_C$ for $\mu_o H = 7$ and 14 T, and increasing H leads to an expansion of the region of $\Delta V/\Delta I = 0$. The critical current $I_C$ as well as $I_{NDR1}$ and $I_{NDR2}$ as a function of **(e)** temperature and **(f)** magnetic field. Note that $I_C$ is depressed with increasing temperature and eventually vanishes near $T_C$ but is enhanced and remains sharp with increasing H || *c*-axis.

**Fig. 4. Time-dependent bistable switching: (a, b, c)** The *a*-axis voltage $V_a$ as a function of time t at T = 10 K for **(a)** H = 0, **(b)** $\mu_o H_{||c} = 7$ T and **(c)** $\mu_o H_{||a} = 7$ T. **(d)** The electric energy W due to applied I as a function of I. The schematics for **(e)** H || *c*-axis, which couples to and enhances $\Psi_C$, corresponding to the case in (b) and **(f)** H || *a*-axis, which suppresses $\Psi_C$, corresponding to the case shown in (c). Note *c*-axis $M_{COC}$ are denoted by the small red and blue arrows in the octahedron.

**Fig. 5. COC below $T_C$** The magnetic symmetry group is *C2'/c'* (No. 15.89, BNS setting) [31], and there are four independent current patterns that preserve the magnetic symmetry group and the crystal space group *P-31c* (No. 163) [28]. **(a)** The four-parameter, ferrimagnetic net, non-zero *c*-axis magnetization. The red, cylindrical arrows are the Mn spins. Four COC are denoted by small

arrows (red, blue, green, and purple). The top view of the Mn1 plane (left) and the side view (right). Details of **(b)** the Mn1 plane and **(c)** the Mn2 plane [28 for more details].

***Methods***

Single crystals of $Mn_3Si_2Te_6$ were grown using a flux method. Measurements of crystal structures were performed using a Bruker Quest ECO single-crystal diffractometer with an Oxford Cryosystem providing sample temperature environments ranging from 80 K to 400 K. Chemical analyses of the samples were performed using a combination of a Hitachi MT3030 Plus Scanning Electron Microscope and an Oxford Energy Dispersive X-Ray Spectroscopy (EDX). The measurements of the electric resistivity and I-V characteristic were carried out using a Quantum Design (QD) Dynacool PPMS system having a 14-Tesla magnet and a set of external Keithley meters that provides current source and measures voltage with a high precision. The magnetization with electric current was measured using a Quantum Design MPMS-XL magnetometer with a homemade probe. The magnetostriction was measured using a home-made dilatometer compatible with the QD Dynacool PPMS system. The dilatometer is made with four identical KYOWA, type KFL strain gauges forming a Wheatstone bridge with the sample mounted on one arm and the rest three as compensators to cancel unwanted changes in the strain gauges due to changes in temperature and/or magnetic field.

***Acknowledgments***

This work is supported by National Science Foundation via Grants No. DMR 1903888 and DMR 2204811. The theoretical part of this work is in part performed at Aspen Center for Physics, which is supported by National Science Foundation grant PHY-1607611. The work at ORNL's SNS is sponsored by the Scientific User Facilities Division, Office of Basic Energy Sciences, U.S. Department of Energy. G.C. thanks Minhyea Lee, Rahul Nandkishore, Xi Chen, Michael Hermele,

David Singh, Dmitry Reznik, Daniel Dessau, and Noel Clark for useful discussions. I.K. thanks Erez Berg, Martin Mourigal, Bruno Uchoa, Chandra Varma, and Ziqiang Wang for useful discussions.

***Author contributions***

Y.Z. conducted measurements of the physical properties and data analysis; Y.F.N. grew the single crystals, characterized the crystal structure of the crystals, measured magnetization with applied currents and contributed to the data analysis; H.D.Z. conducted measurements of crystal and physical properties and data analysis; F.Y. determined the magnetic structure of the crystals using neutron diffraction and contributed to the data analysis; S.H. contributed to the theoretical analysis including detailed configurations of chiral orbital currents presented in the main text and Supplemental Material; L.E.D. contributed to the data analysis and paper writing. I.K. proposed the theoretical argument, formed the theoretical discussion of the paper and contributed to paper writing; G.C. initiated and directed the work, analyzed the data and wrote the paper.

***Competing interests***

None.

***Materials & Correspondence***

Professor Gang Cao, gang.cao@colorado.edu

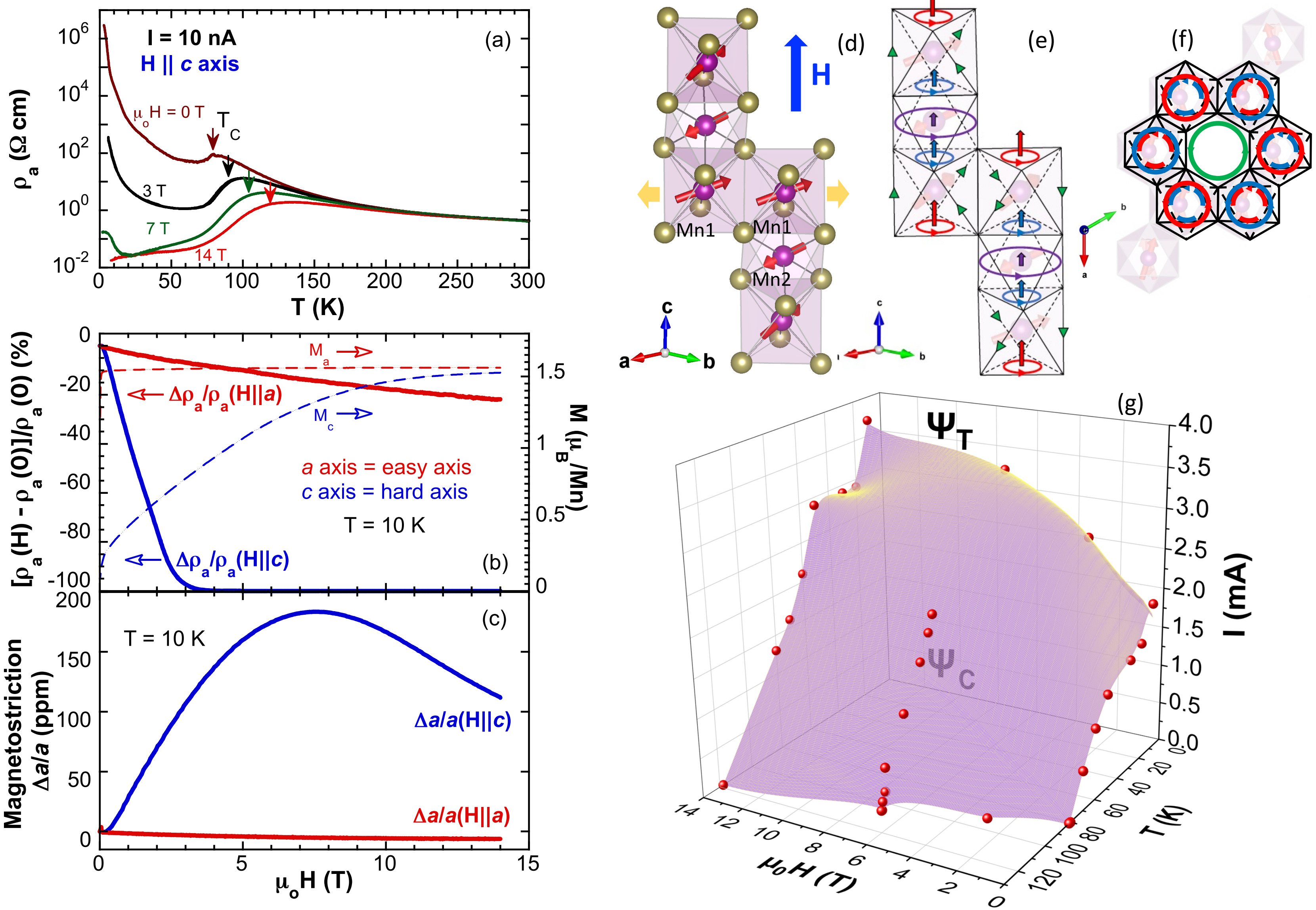

Figure 1

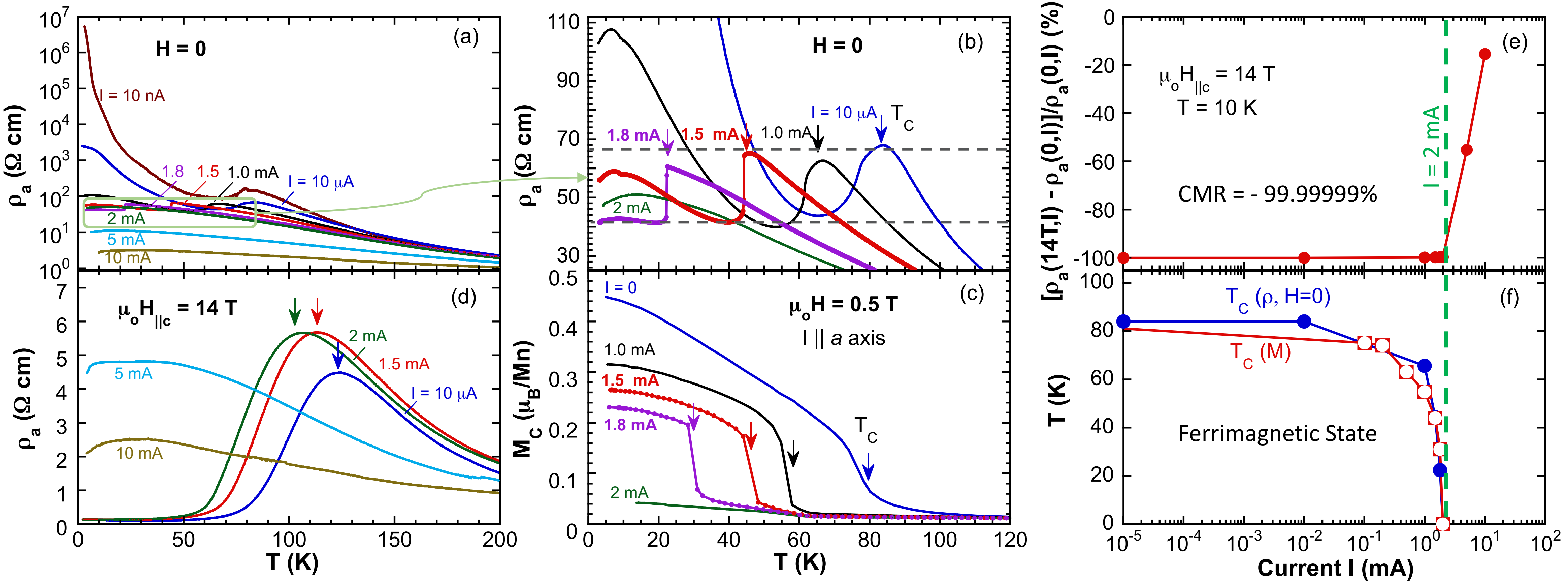

H = 0
(a)
I = 10 nA
1.8
1.5
1.0 mA
I = 10 μA
2 mA
5 mA
10 mA
ρa (Ω cm)
H = 0
(b)
1.8 mA
1.5 mA
1.0 mA
I = 10 μA
TC
2 mA
(e)
μoH||c = 14 T
T = 10 K
CMR = - 99.99999%
I = 2 mA
[ρa(14T,I) - ρa(0,I)]/ρa(0,I) (%)
μoH||c = 14 T
(d)
2 mA
1.5 mA
I = 10 μA
5 mA
10 mA
T (K)
I = 0
μoH = 0.5 T
I || a axis
(c)
1.0 mA
1.5 mA
1.8 mA
TC
2 mA
MC (μB/Mn)
TC (ρ, H=0)
(f)
TC (M)
Ferrimagnetic State
T (K)
Current I (mA)

Figure 2

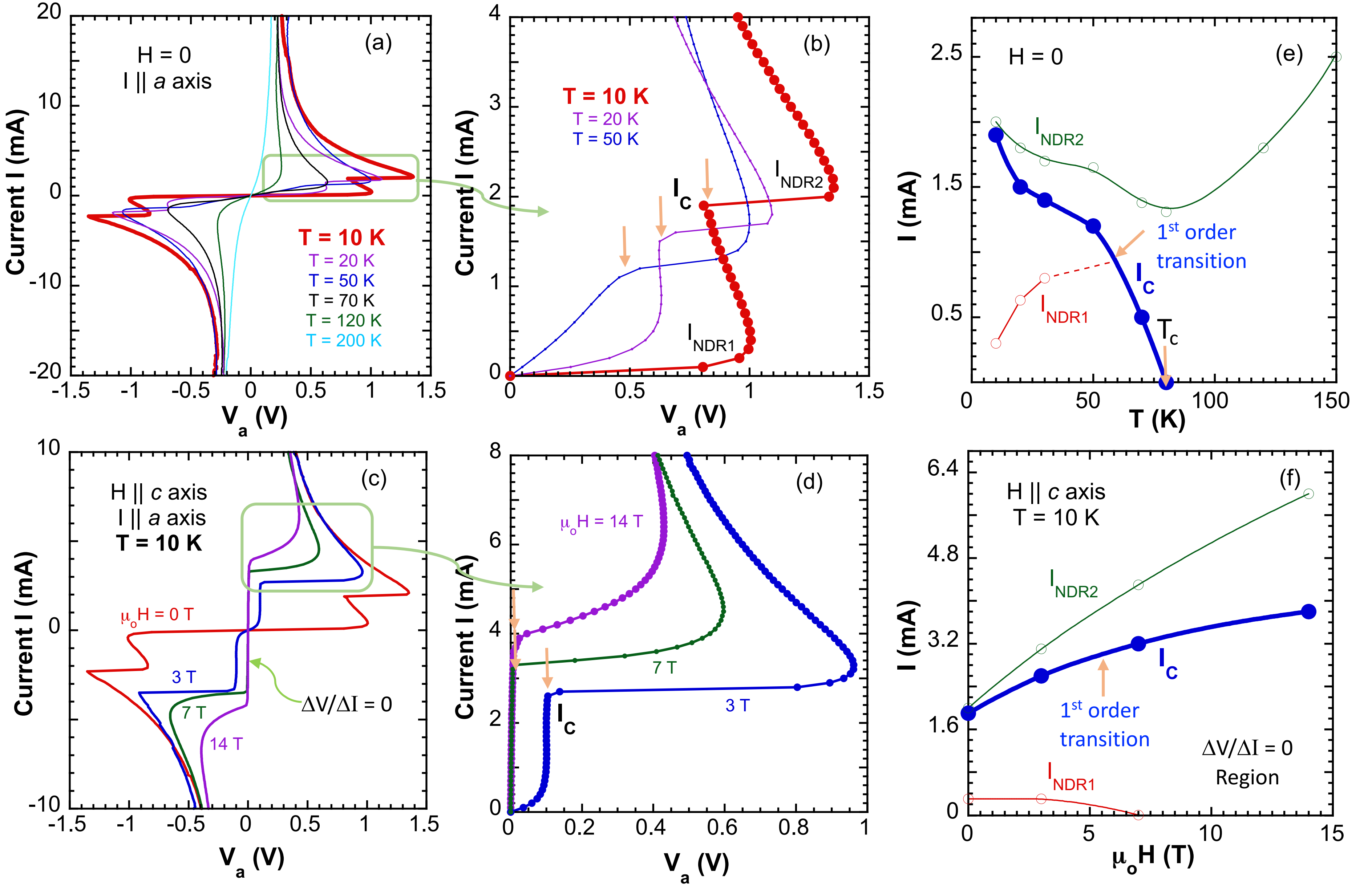
H = 0
I || a axis
(a)
T = 10 K
T = 20 K
T = 50 K
T = 70 K
T = 120 K
T = 200 K
Current I (mA)
V_a (V)
(b)
T = 10 K
T = 20 K
T = 50 K
I_C
I_NDR1
I_NDR2
Current I (mA)
V_a (V)
H = 0
(e)
I_NDR2
1st order transition
I_C
I_NDR1
T_C
I (mA)
T (K)
H || c axis
I || a axis
T = 10 K
(c)
μoH = 0 T
3 T
7 T
14 T
ΔV/ΔI = 0
Current I (mA)
V_a (V)
(d)
μoH = 14 T
7 T
3 T
I_C
Current I (mA)
V_a (V)
H || c axis
T = 10 K
(f)
I_NDR2
I_C
1st order transition
ΔV/ΔI = 0
Region
I_NDR1
I (mA)
μoH (T)

Figure 3

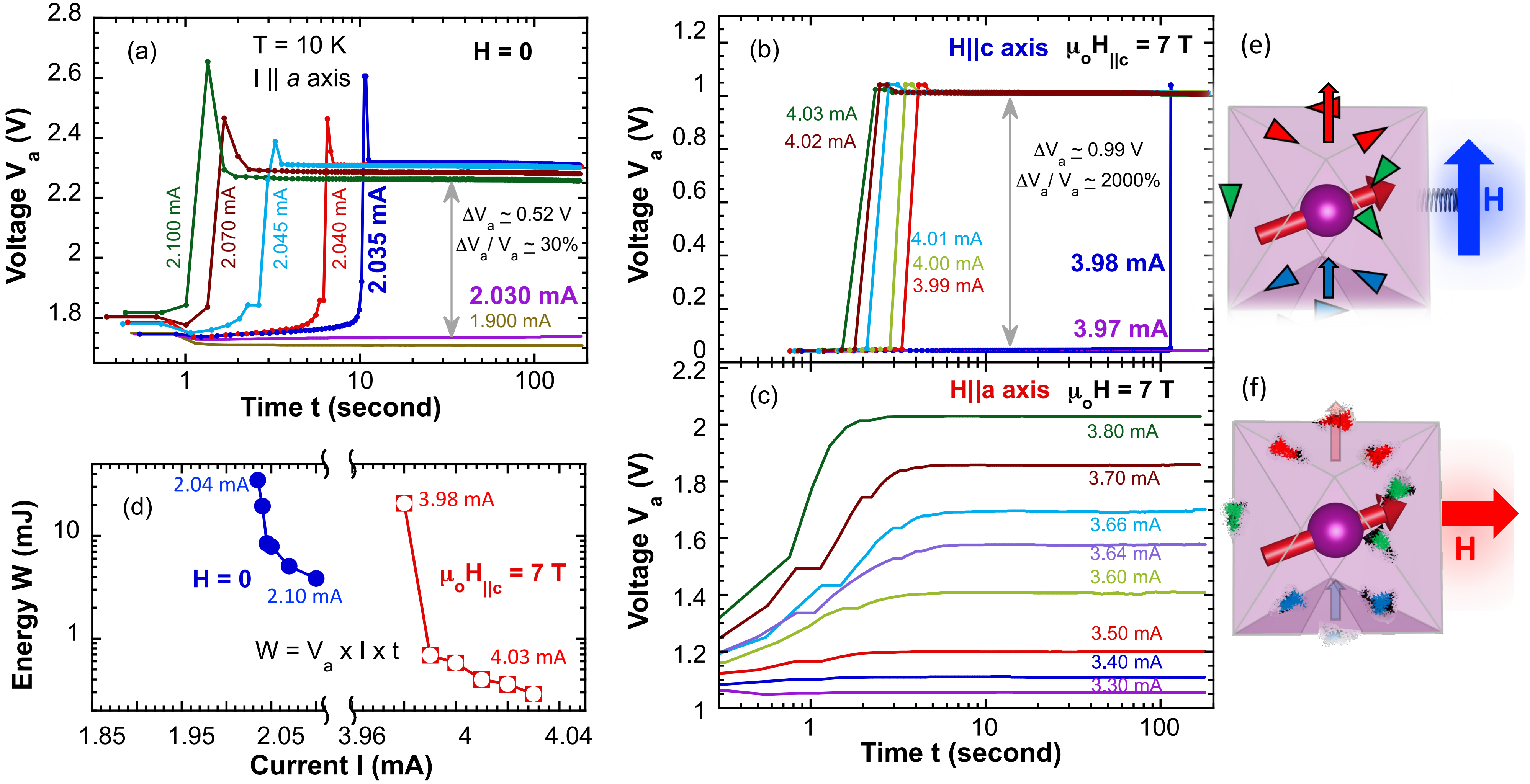
(a)
T = 10 K
I || a axis
H = 0
2.100 mA
2.070 mA
2.045 mA
2.040 mA
2.035 mA
2.030 mA
1.900 mA
ΔV_a ≃ 0.52 V
ΔV_a / V_a ≃ 30%
Voltage V_a (V)
Time t (second)
(b)
H||c axis μ_oH_||c = 7 T
4.03 mA
4.02 mA
4.01 mA
4.00 mA
3.99 mA
3.98 mA
3.97 mA
ΔV_a ≃ 0.99 V
ΔV_a / V_a ≃ 2000%
(c)
H||a axis μ_oH = 7 T
3.80 mA
3.70 mA
3.66 mA
3.64 mA
3.60 mA
3.50 mA
3.40 mA
3.30 mA
(d)
2.04 mA
H = 0
2.10 mA
3.98 mA
μ_oH_||c = 7 T
4.03 mA
W = V_a x I x t
Energy W (mJ)
Current I (mA)
(e)
(f)
H

Figure 4

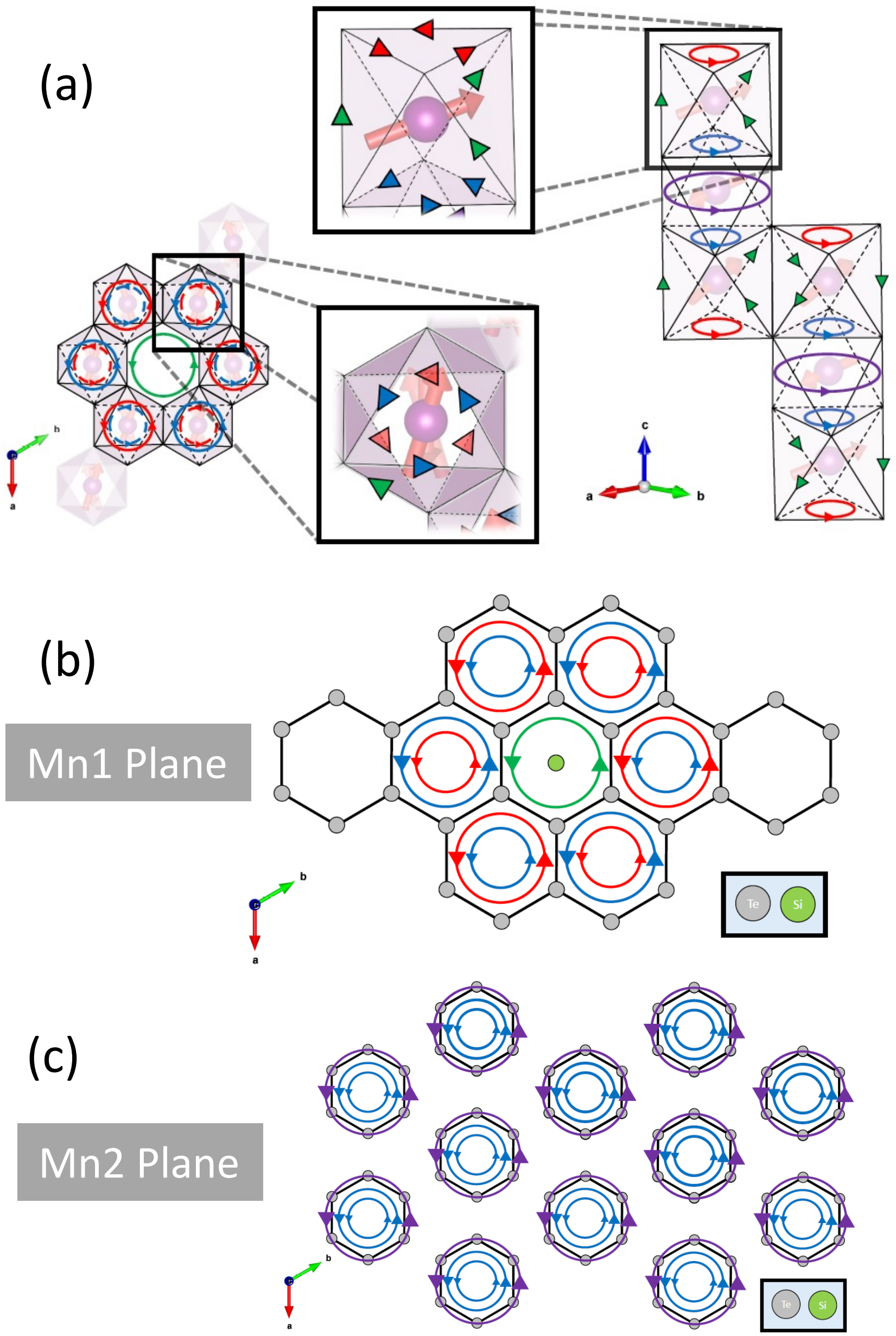

Figure 5

SUPPLEMENTAL MATERIAL

# Control of chiral orbital currents in a colossal magnetoresistance material

Yu Zhang[1], Yifei Ni[1], Hengdi Zhao[1], Sami Hakani[2], Feng Ye[3], Lance DeLong[4], Itamar Kimchi[2] and Gang Cao[1]

[1]*Department of Physics, University of Colorado at Boulder, Boulder, CO 80309, USA*

[2]*School of Physics, Georgia Institute of Technology, Atlanta, GA 30332, USA*

[3]*Neutron Scattering Division, Oak Ridge National Lab, Oak Ridge, Tennessee 37831, USA*

[4]*Department of Physics and Astronomy, University of Kentucky, Lexington KY 40506, USA*

## I. Additional notes on the chiral orbital currents, colossal magnetoresistance, and bistable switching in $Mn_3Si_2Te_6$

1. The chiral orbital currents (COC) reported in the main text not only coexist with long range magnetic order but also produce their own nonzero net magnetization. The COC discussed in the cuprates and other materials involve zero net circulation, and zero net magnetic moment; By their observed symmetries, they necessarily entail an "antiferromagnetic" pattern of clockwise and counterclockwise COC circulations. This distinction is key to what makes the COC of this present material unique.
2. It is important to note that colossal magnetoresistance (CMR) is in the lowest order tensor, unlike high-order, non-linear electric response known as electronic magneto-chiral anisotropy (eMChA).
3. We would like to point out that our picture of COC requires strongly correlated physics that are irreconcilable with a picture purely based on electronic bands with a band gap [1]. Our model goes beyond simple modifications of a band structure due to interactions.
4. The CMR is enhanced (**SM-Fig.1**), while the saturated magnetization $M_s$ is reduced to 1.40 $\mu_B$/Mn in $Mn_3(Si_{1-x}Ge_x)_2Te_6$, which is further evidence that larger orbital moments of the COC state partially cancel the net magnetic moments. On the other hand, in $Mn_3Si_2(Te_{1-y}Se_y)_6$ the CMR is weakened (**SM-Fig.1**), but $M_s$ is enhanced up to 1.7 $\mu_B$/Mn, which indicates smaller orbital moments that lie on the mixed-occupancy, chalcogenide sublattice have a weaker cancellation effect.

It is also noteworthy that Ge doping that enhances the CMR also expands the *ab*-plane, whereas Se doping that weakens the CMR shrinks the *ab*-plane (Inset in **SM-Fig.1**).

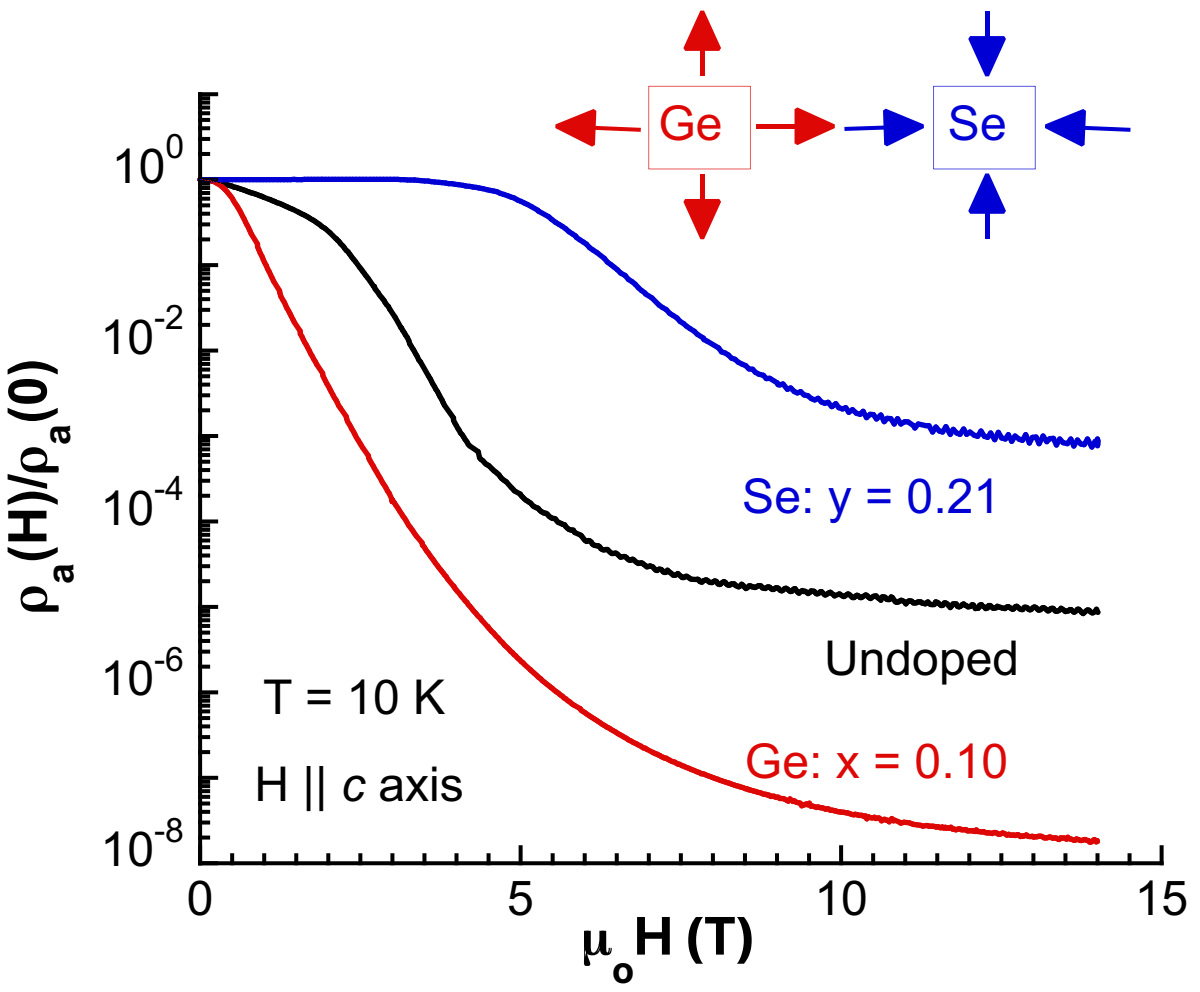

**SM-Fig.1.** The magnetic field dependence of the *a*-axis magnetoresistance ratio defined by $\rho_a(H)/\rho_a(0)$ for $Mn_3(Si_{1-x}Ge_x)_2Te_6$ (red), $Mn_3Si_2(Te_{1-y}Se_y)_6$ (blue) and undoped compound (black). Inset: schematic illustration of the unit cell expansion and contraction due to Ge doping (red) and Se doping (blue), respectively

5. The orbital magnetic moments due to the COC are estimated based on the total measured moments, 4.55 and 4.20 Bohr magneton for Mn1 and Mn2, respectively [2], which are significantly smaller than expected 5 Bohr magneton. The reduced magnetic moments suggest partial cancellation by the COC orbital magnetic moments. We infer that the upper limit of the COC orbital magnetic moments is on the order of 0.10 Bohr magneton. We are mindful that it is always possible that the Mn local moments could be smaller than expected, thus less cancellation with the COC orbital magnetic moments would be needed to produce the total observed moments.

6. The magnetostriction is associated with the COC orbital moments because the Lorentz force acts to expand an *ab*-plane COC circulating on the rigid Te-Te edges of $MnTe_6$ octahedra when H || *c*-axis. The net orbital magnetization, which equals the loop current times the orbital area, naturally increases due to increasing the orbital area.
7. The S-shaped negative differential resistance is uncommon in bulk materials, it highly desirable for nonvolatile memory devices [3, 4].
8. The rigid coupling of the COC to the Te sublattice and $M_{Mn}$ is a signature of the new state. The COC extend over multiple atomic sites, with couplings that depend on the ionic positions and dictate the current strength. The strong coupling sets time scales, which enables the COC state to remain metastable over seconds or minutes before undergoing a first-order transition to a trivial state when applied DC currents exceed certain critical values (on the order of 1 mA). Such a long-time scale is rare, if not unprecedented, but understandable. In the main text, we infer that this transition may mimic an ice-to-water phase transition. During ice melting, the thermal energy is spent to break the stiff hydrogen bonds, without raising temperature. The temperature of the system rises abruptly only when all hydrogen bonds are broken, i.e., the ice is completely melted. An analogy drawn here is that at the metastable state or during the COC melting, the applied DC currents (on the order of 1 mA) circulate in the sample to break COC in every unit cell, without causing a voltage or resistance increase before all COC are completely melted. This is because electrons always flow along the least resistive path, and this case, the remaining COC, according to the two-channel model. Breaking the COC inevitably causes rearrangements of Te orbitals and atoms, more generally, changing lattice properties, therefore a long delay for switching. The larger the applied currents, the faster they can destroy all COC, thus a shorter time delay.

## II. Experimental details

Single crystals of $Mn_3Si_2Te_6$ were grown using a flux method [29]. Measurements of crystal structures were performed using a Bruker Quest ECO single-crystal diffractometer with an Oxford Cryosystem providing sample temperature environments ranging from 80 K to 400 K. Chemical analyses of the samples were performed using a combination of a Hitachi MT3030 Plus Scanning Electron Microscope and an Oxford Energy Dispersive X-Ray Spectroscopy (EDX). The measurements of the electric resistivity and I-V characteristic were carried out using a Quantum Design (QD) Dynacool PPMS system having a 14-Tesla magnet and a set of external Keithley meters that provides current source and measures voltage with a high precision. The magnetization with electric current was measured using a Quantum Design MPMS-XL magnetometer with a homemade probe. The magnetostriction was measured using a home-made dilatometer compatible with the QD Dynacool PPMS system. The dilatometer is made with four identical KYOWA, type KFL strain gauges forming a Wheatstone bridge with the sample mounted on one arm and the rest three as compensators to cancel unwanted changes in the strain gauges due to changes in temperature and/or magnetic field.

## III. Additional data

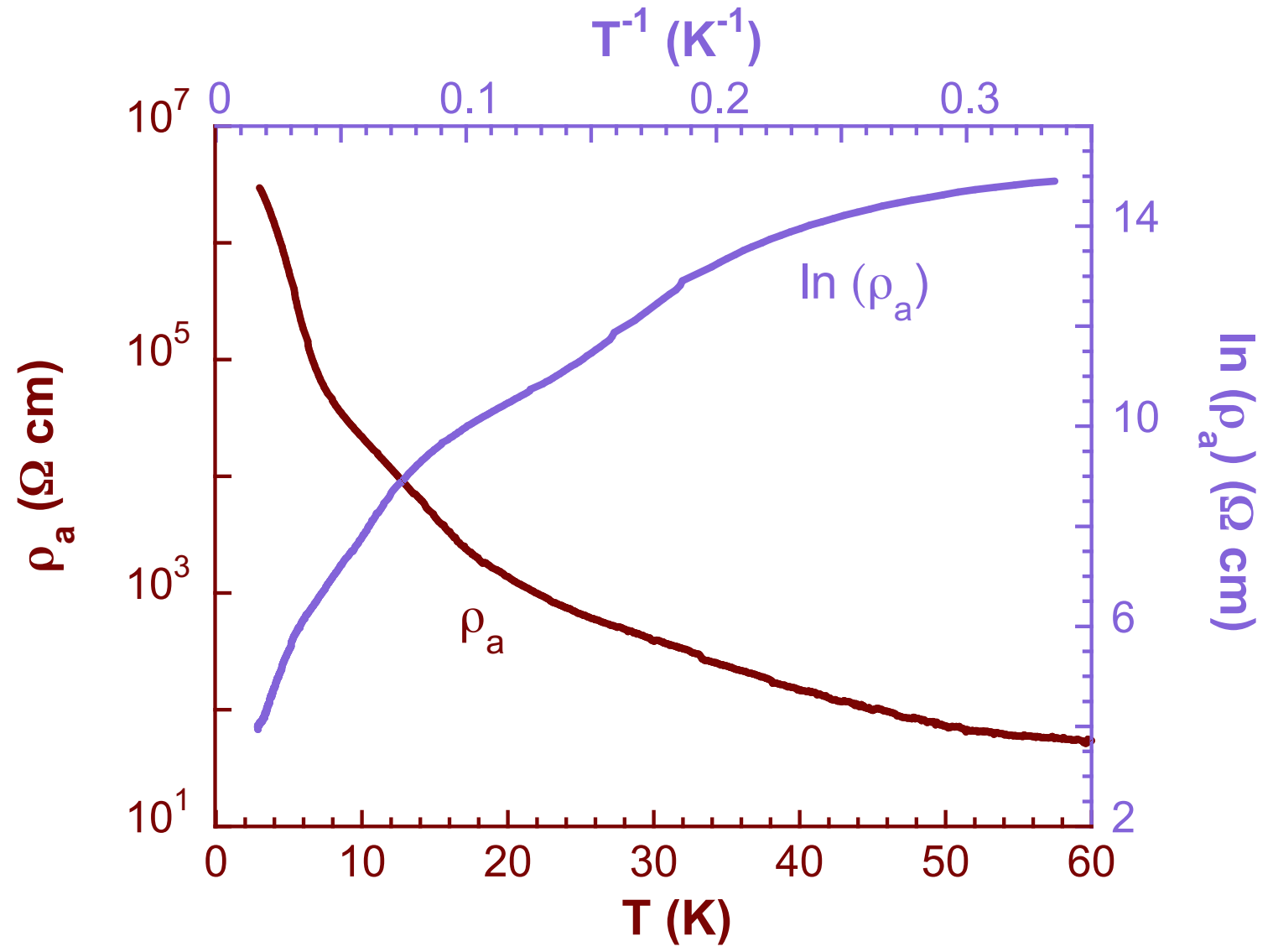

**SM-Fig.2. Resistivity at low temperatures:** The temperature dependence of the *a*-axis resistivity $\rho_a$ at low temperatures (data in brown), and ln ($\rho_a$) vs. $T^{-1}$ (data in light purple). Note that $\rho_a$ does not follow an activation law and or a simple power law at low temperatures.

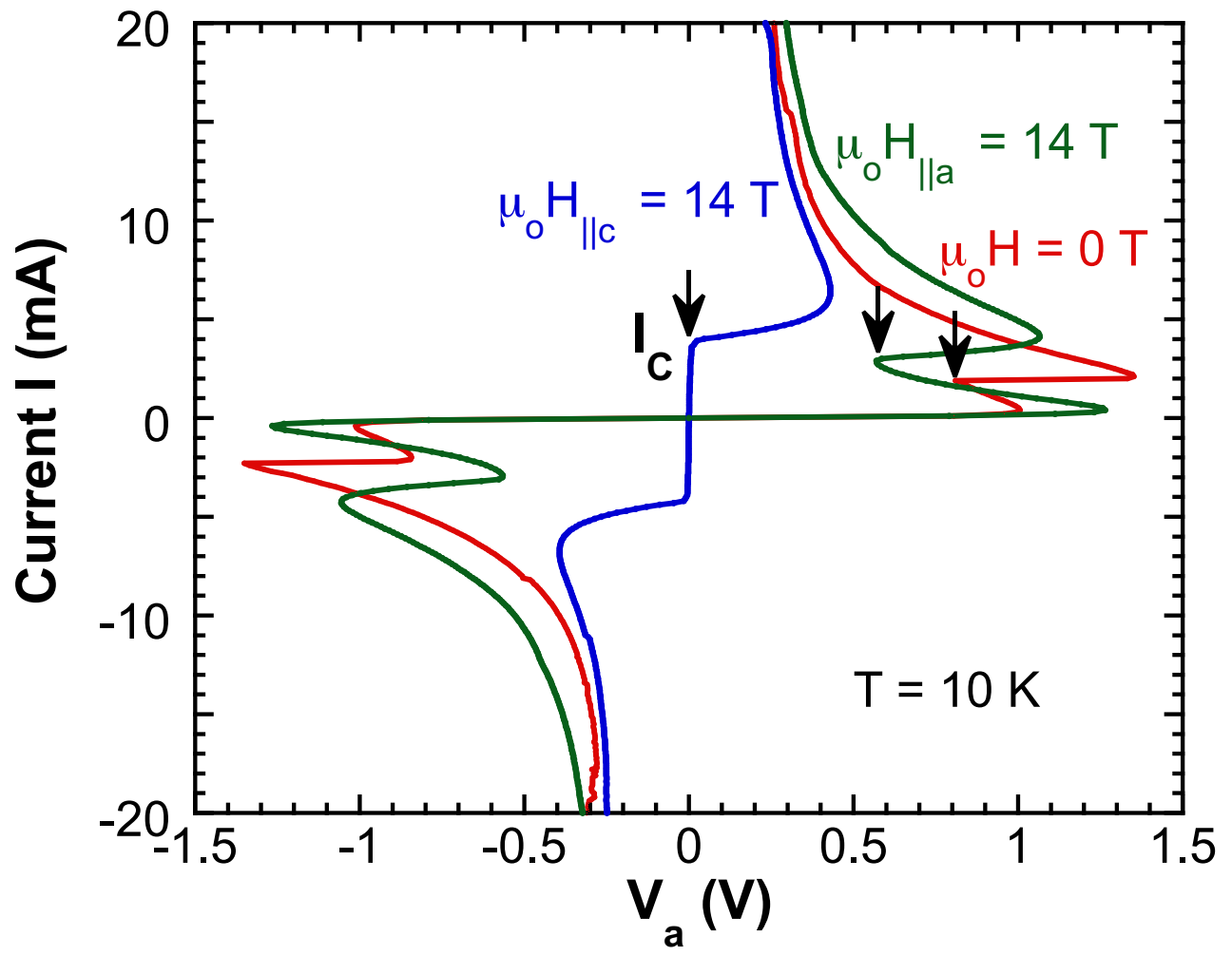

**SM-Fig.3. Comparison of the I-V characteristic at H || *a*-axis and H || *c*-axis:** The *a*-axis I-V characteristic at 10 K for H = 0 (red), $\mu_o H$ = 14 T along the *a*-axis (green) and the *c*-axis (green). Note the regime where $\Delta V/\Delta I = 0$ emerges only when H || *c*-axis.

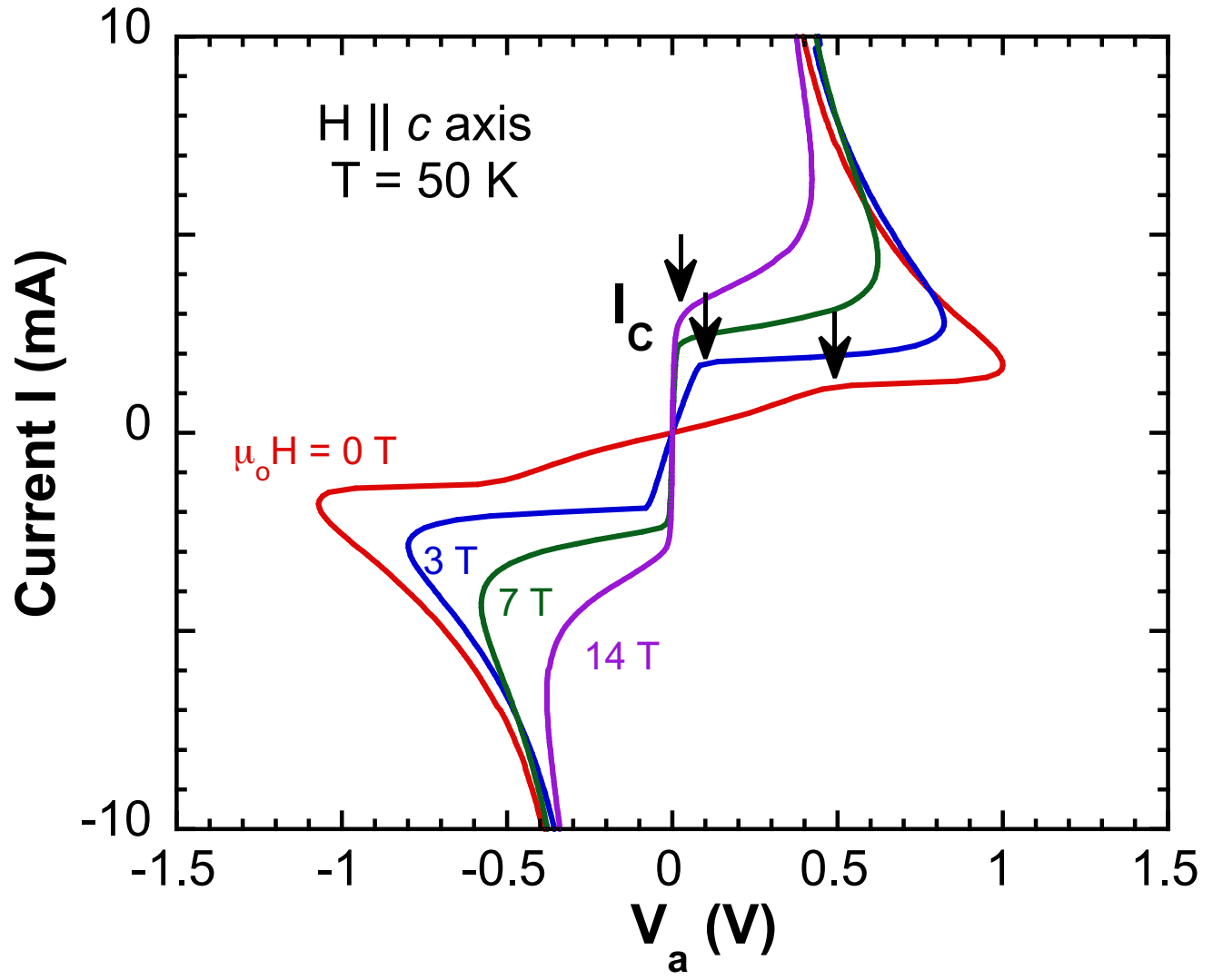

**SM-Fig.4. The I-V characteristic at H || *c*-axis and 50 K:** Note the regime $\Delta V/\Delta I = 0$ persists at 7 T and 14 T.

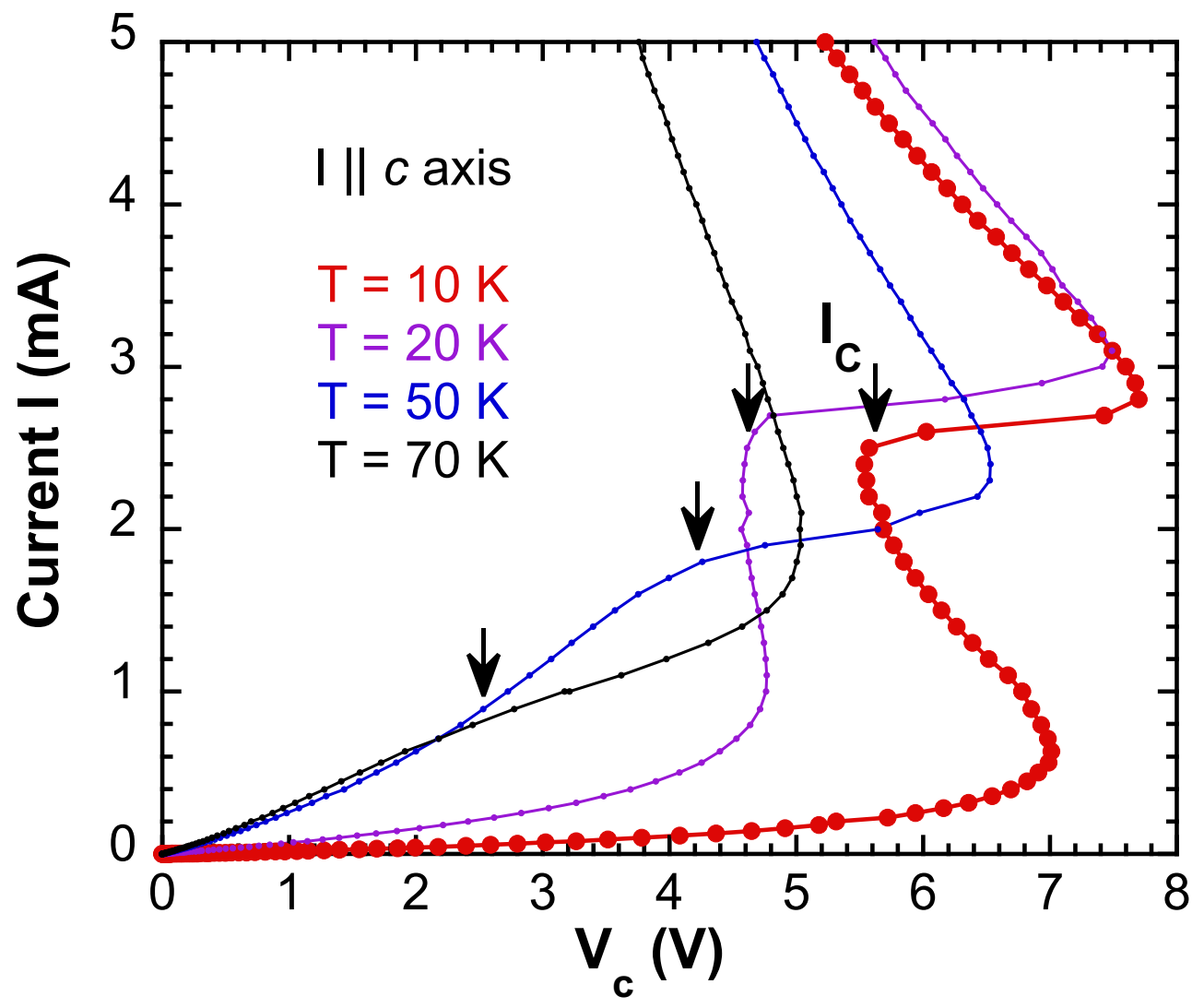

**SM-Fig.5. The I-V characteristic at I || *c*-axis for various temperatures:** Note that the I-V characteristic is qualitatively similar to that for I || *a*-axis but the first-order switching at $I_C$ is weaker.

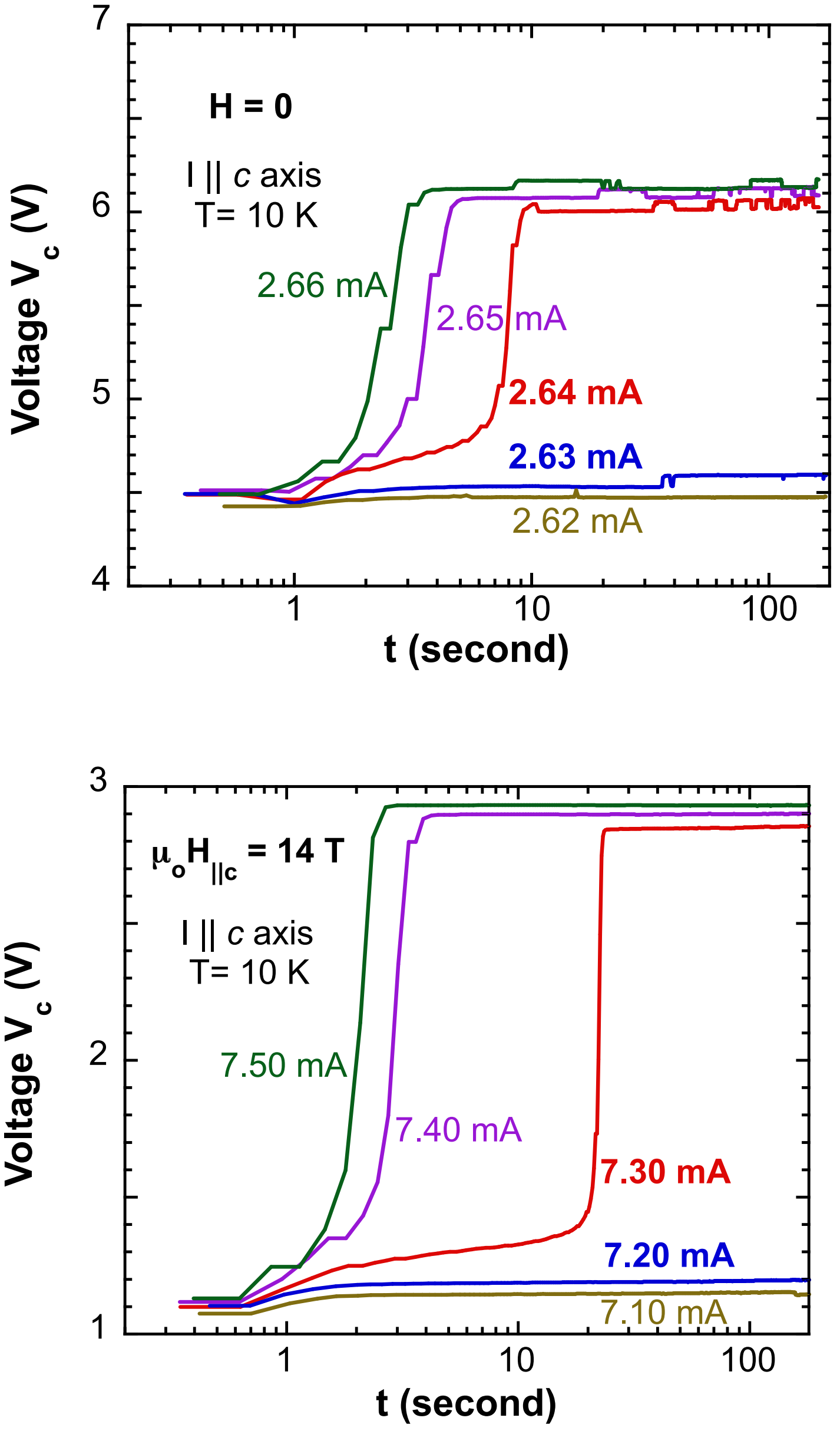

**SM-Fig.6. Time-dependent bistable switching for current applied along the *c*-axis at 10 K for**

H = 0 (upper panel) and $\mu_o$H = 14 T (lower panel).

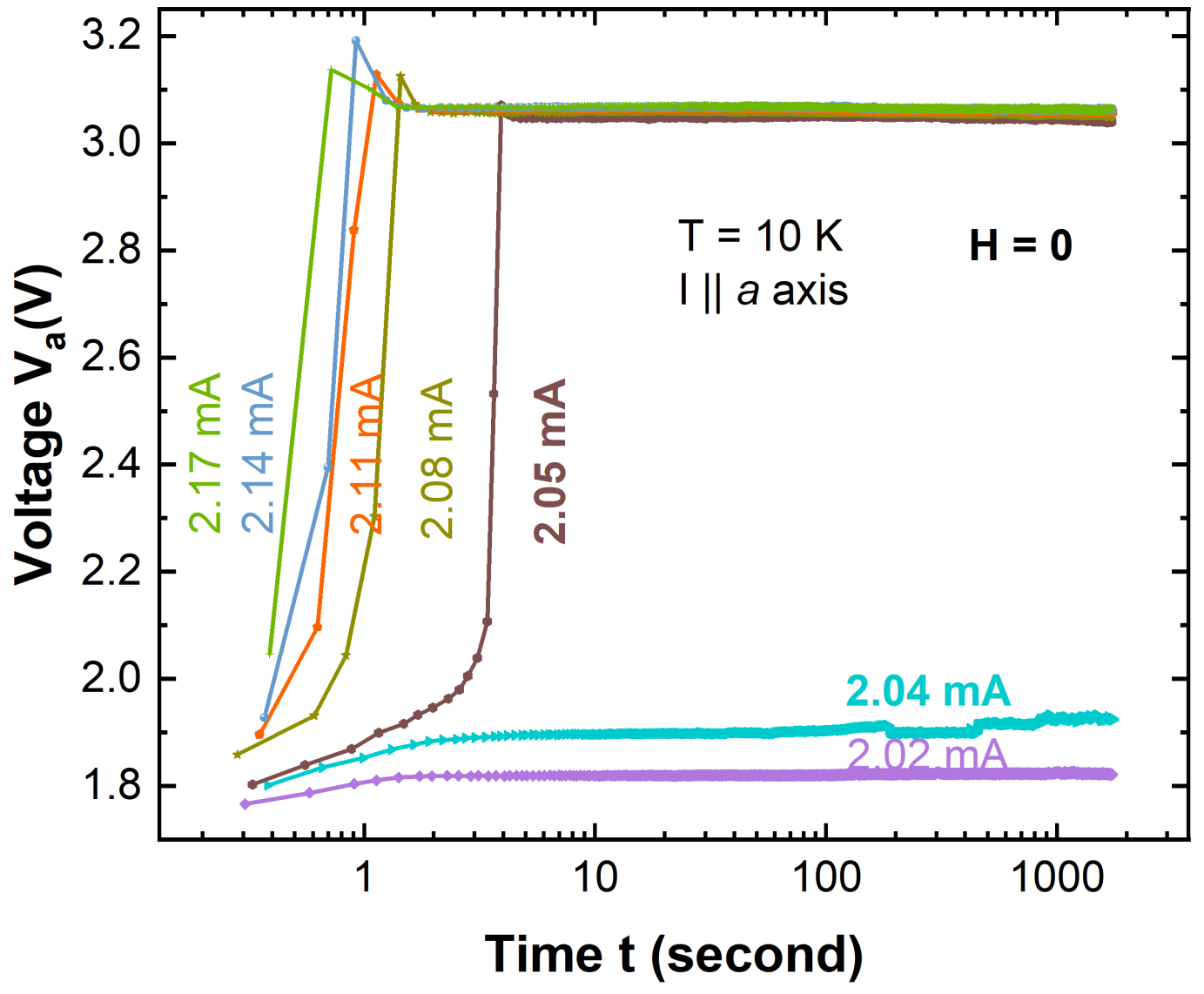

**SM-Fig.7. Time-dependent bistable switching at 10 K with 1,800 seconds elapsed.**

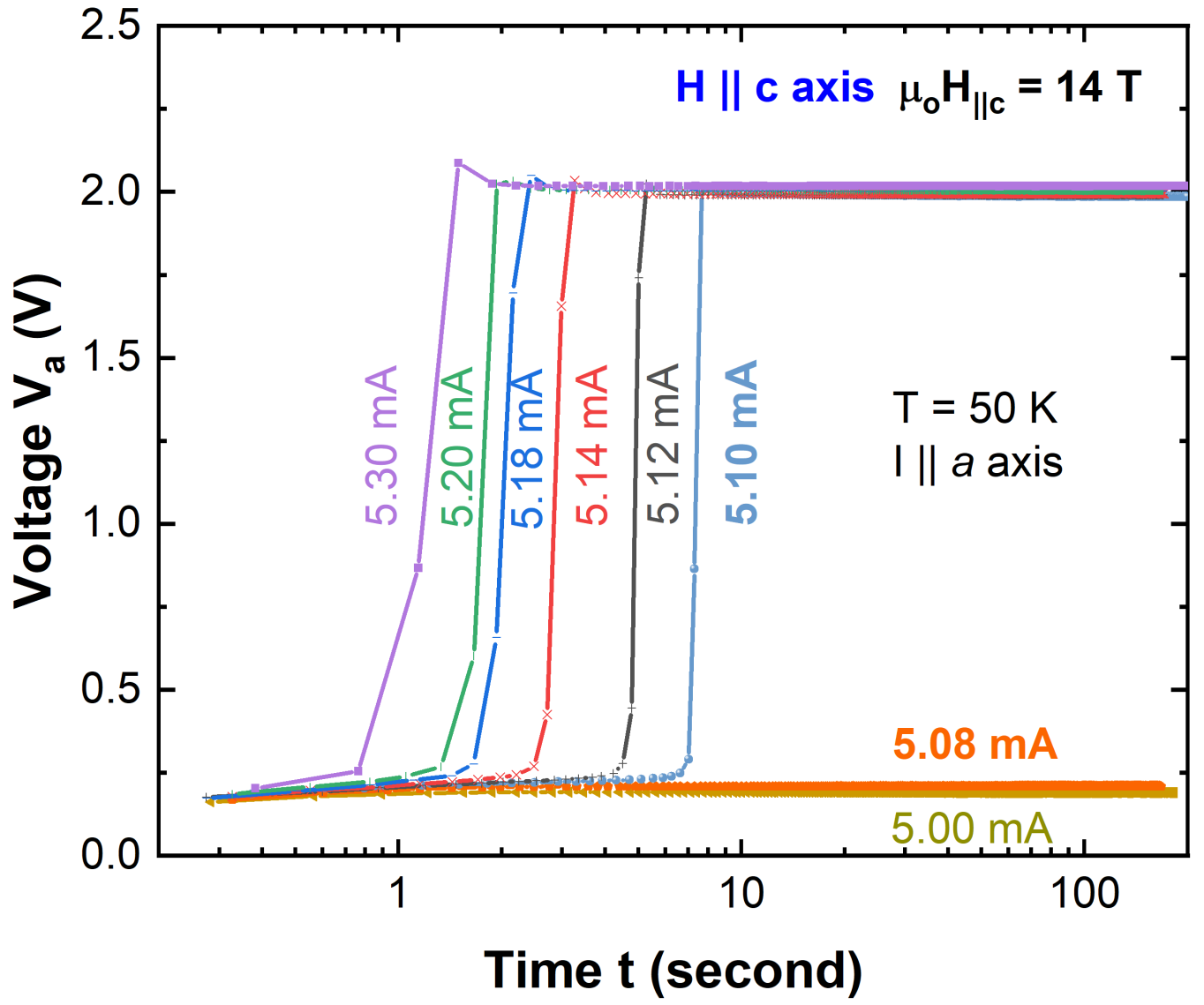

**SM-Fig.8. Time-dependent bistable switching at 50 K and $\mu_o H_{||c}$ = 14 T**

## IV. Additional discussion on chiral orbital currents

Below $T_C$, the ferrimagnetic order is observed with magnetic symmetry group *C2'/c'* (No. 15.89, BNS setting) [2]. This symmetry group allows certain configurations of Te orbital currents circulating within the unit cell. Since we are particularly interested in the COC configurations with *c*-axis orbital moments, we focus on one particular subset of COC patterns that respect not only the magnetic group 15.89, but also the additional 3-fold rotational symmetry inherited from the full crystal space group *P-31c* (No. 163). There are four independent current patterns that preserve this combination of 15.89 plus 3-fold rotations. The resulting four-parameter COC state is depicted in **Figs. 5a-5c** in the main text.

This symmetry-preserving COC state $\Psi_C$ yields currents circulating on octahedral top/bottom faces as well as around Mn2 sites and around the Si sites. The resulting magnetic moments are all oriented exactly along the *c*-axis and can produce a nonzero net *c*-axis moment. There are four additional independent current patterns that preserve the magnetic group 15.89 but break 3-fold rotation symmetry; these patterns, denoted $\Psi_a$, are described below, and produce moments oriented within the *ab*-plane, coupling only weakly to *c*-axis fields. Henceforth we focus on $\Psi_C$, which gives the primary contribution to the COC state and its response to *c*-axis fields.

The COC of $\Psi_C$ circulate along Te-Te octahedra edges and primarily involve Te orbitals. The Mn local moment pattern detected in neutron scattering [2] linearly couples to the order parameter of $\Psi_C$ in a Landau-Ginzburg theory. Therefore, the orbital currents and localized moments both contribute to the observed magnetic pattern and cannot be fully distinguished. As such, $\Psi_C$ can be viewed as a superposition of Mn and Te states that forms time-reversal-breaking multi-site molecular orbitals.

Application of H || *c*-axis couples to $\Psi_C$ via the *c*-axis orbital moments, which in turn drastically enhances the COC order parameter, therefore the hopping matrix elements, and accommodates more current passing through the sample (**Fig. 5a** in the main text), leading to the observed CMR. Clearly application of H || *a*-axis weakens the *c*-axis orbital moments, thus reducing $\Psi_C$, and necessarily causes suppression of the CMR. On the other hand, applying uniform current I > 2mA (along the *a* or *c* axes) weakens the COC order parameter, therefore destabilizes $\Psi_C$ in favor of the trivial state $\Psi_T$. This is expected since the COC configuration does not permit any net current because each of the four independent parameters that specify $\Psi_C$ forms a circulating current within the unit cell with no net uniform component (**Figs. 5a-5c** in the main text). The resistivity is reduced with increased I; but at the same time, the increased I also weakens and eventually suppresses the ferrimagnetic state, and consequently $\Psi_C$, leading to a first-order phase transition that destroys the COC state (**Figs. 2a-2d** in the main text).

Strong electron-lattice coupling is an important feature of $\Psi_C$. The COC extend over multiple atomic sites, with couplings that depend on the ionic positions in the lattice and determine the current strengths. Therefore, $\Psi_C$ entails a much stronger coupling between the lattice and the *c*-axis magnetic moments, and associated anisotropic magnetoelastic coupling, compared to that due to conventional $Mn^{+2}$-site, spin-only moments. This property of $\Psi_C$ is consistent with the large magnetostriction $\Delta a/a$ when H || *c*-axis and the nearly negligible $\Delta a/a$ when H || *a*-axis (**Fig. 1c in** the main text). Note that these effects occur only below $T_C$ where $\Psi_C$ exists.

The strong coupling of COC to the lattice suggests that the current-driven transition from the COC to the trivial state should be strongly first order as observed (**Figs. 1-4** in the main text). The long-time scale (seconds or minutes) for the bistable switching is consistent with a picture wherein melting $\Psi_C$ involves changing lattice properties (e.g., bond lengths or fluctuations/phonon effects), thus allowing $\Psi_C$ to remain metastable over long time scales set by *ionic motion* rather than just by electrons and/or Mn spins. The coupling between $\Psi_C$ and the lattice increases when H || *c*-axis, thus the observed longer time delay, but significantly weakened when H || *a*-axis, leading to a mixed state with coexisting $\Psi_C$ and $\Psi_a$.

**V. Additional symmetry discussion**

As depicted in **Fig. 5** of the main text, $\Psi_C$ respects both (1) a lower monoclinic magnetic space group *C2'/c'* (No. 15.89, BNS setting) which correctly describes the canted magnetic structure [26], and (2) an extra 3-fold rotational symmetry about the *c*-axis endowed from the full crystal space group *P-31c* (No. 163). Additional COC are symmetry allowed by (1), but they do not couple as strongly to applied fields in the *c*-axis. These additional COC are shown in **SM-Fig. 9** and parametrize the state $\Psi_a$.

The full COC state is parametrized by eight independent COC. These include three of the COC of $\Psi_C$ which have *c*-axis moments (red, blue, green in **Fig. 5** in the main text) and the six COC shown in **SM-Fig. 9** which do not constitute a linearly independent set of COC. One of COC (purple in **Fig. 5**) is in fact a linear combination of COC that circulate around Mn1 atoms (**SM-Fig. 9** (left)) and different COC from $\Psi_C$ (blue in **Fig. 5**).

The sum of the three COC circulating about Mn1 atoms (**SM-Fig. 9** (left)) in equal magnitudes in fact gives the difference of two COC of $\Psi_C$ (red minus blue in **Fig. 5** in the main text). Moreover, the sum of the three COC in equal magnitudes circulating about the Mn2 atoms (**SM-Fig. 9** (right)) gives another difference of two COC of $\Psi_C$ (blue minus twice purple). These linear combinations are symmetry allowed by the 3-fold rotational symmetry described above. In total, the COC state thus has the four parameters of $\Psi_C$, six additional parameters symmetry allowed by the magnetic space group, and two constraints relating the ten parameters. Indeed, the full COC state is thus parametrized by eight independent COC.

The unit vector normal to the plane about which each COC circulates in $\Psi_C$ is exactly along the *c*-axis. The COC of this state clearly couple most strongly to applied fields in the *c*-axis. In contrast, the COC of $\Psi_a$ do not couple strongly to applied fields in the *c*-axis. Instead, the orbital moments generated by the COC in the Mn1 plane point primarily in the *ab*-plane, and they are (approximately) pairwise separated by 120°. Their orientation is slightly nontrivial since the $MnTe_6$ octahedra are not regular octahedra. The orbital moments generated by the COC in the Mn2 plane similarly are not well aligned with the *c*-axis, and their orientation is also nontrivial.

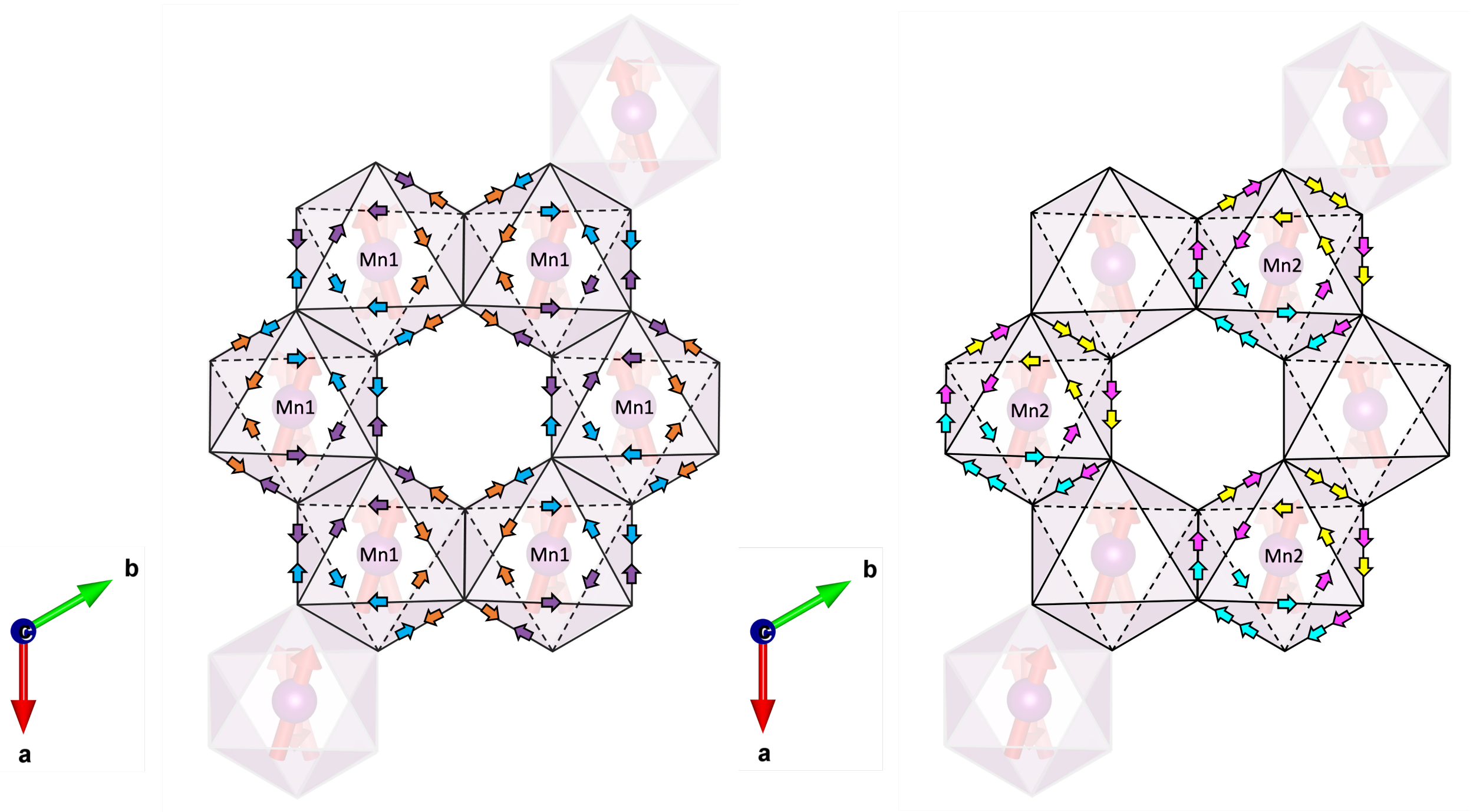

**SM-Fig. 9. Chiral orbital current parameters of $\Psi_a$ in the Mn1 (left) and Mn2 (right) planes:** Three independent currents (orange, purple, and cerulean) run along Te-Te bonds in the Mn1 plane (left) and are symmetry allowed magnetic space group. Three more independent currents (cyan, magenta, and yellow) run along Te-Te bonds in the Mn2 plane (right). These currents are not linearly independent of the COC of **Fig. 5** of the main text. Explicitly, the sum of the orange, purple, and cerulean COC (left) gives the difference of the red and blue COC of **Fig. 5** in the main text. Moreover, the sum of the cyan, magenta, and yellow COC (right) gives the difference of the blue and twice purple COC of **Fig. 5**. Bonds with two arrows of the same color indicate that the current magnitude is doubled on that edge. In total, the COC state is parametrized by eight independent loop currents.